\documentclass[11pt,a4paper]{article}
\usepackage[hyperref]{acl2020}
\usepackage{times}
\usepackage{latexsym}

\usepackage{url}

\usepackage{amsmath,amsfonts,bm}

\def\eqref#1{equation~\ref{#1}}

\def\1{\bm{1}}

\DeclareMathAlphabet{\mathsfit}{\encodingdefault}{\sfdefault}{m}{sl}
\SetMathAlphabet{\mathsfit}{bold}{\encodingdefault}{\sfdefault}{bx}{n}

\usepackage{url}
\usepackage{graphicx}               
\usepackage{tabularx}               
\newcolumntype{C}{>{\centering\arraybackslash}X}
\usepackage{multirow}               
\usepackage{diagbox}                
\usepackage{hhline}                 
\usepackage{color}                  
\usepackage{amsmath}                
\usepackage{amssymb}                
\usepackage{mathtools}              
\usepackage{comment}              

\newcommand{\bs}{\boldsymbol}

\newcommand{\mmee}{M\textsuperscript{2}E\textsuperscript{2}}
\newcommand{\wase}{\textbf{W}eakly \textbf{A}ligned \textbf{S}tructured \textbf{E}mbedding}
\aclfinalcopy 

\usepackage{wrapfig}

\definecolor{RoseQuartzBg}{HTML}{F7CAC9}
\definecolor{RoseQuartz}{HTML}{F5A798}
\definecolor{Serenity}{HTML}{92A8D1}

\usepackage{enumitem}

\usepackage{arydshln}               
\definecolor{themered}{HTML}{FF8375}

\usepackage{xparse}
\NewDocumentCommand{\heng}{ mO{} }{\textcolor{RoseQuartz}{\textsuperscript{\textit{Heng}}\textsf{\textbf{\small[#1]}}}}
\NewDocumentCommand{\shifu}{ mO{} }{\textcolor{RoseQuartz}{\textsuperscript{\textit{Shi-Fu}}\textsf{\textbf{\small[#1]}}}}
\NewDocumentCommand{\manling}{ mO{} }{\textcolor{Serenity}{\textsuperscript{\textit{Manling}}\textsf{\textbf{\small[#1]}}}}
\NewDocumentCommand{\ali}{ mO{} }{\textcolor{Serenity}{\textsuperscript{\textit{Ali}}\textsf{\textbf{\small[#1]}}}}
\NewDocumentCommand{\spencer}{ mO{} }{\textcolor{Serenity}{\textsuperscript{\textit{Spencer}}\textsf{\textbf{\small[#1]}}}}
\NewDocumentCommand{\vicki}{ mO{} }{\textcolor{Serenity}{\textsuperscript{\textit{Qi}}\textsf{\textbf{\small[#1]}}}}
\NewDocumentCommand{\di}{ mO{} }{\textcolor{Serenity}{\textsuperscript{\textit{Di}}\textsf{\textbf{\small[#1]}}}}

\NewDocumentCommand{\method}{ mO{} }{#1}

\usepackage{fixltx2e}
\usepackage{dsfont}

\title{Cross-media Structured Common Space for Multimedia Event Extraction}

 \author{Manling Li\textsuperscript{\textnormal{1}}\thanks{\; These authors contributed equally to this work. },
 Alireza Zareian\textsuperscript{\textnormal{2}}\footnotemark[1],
 Qi Zeng\textsuperscript{\textnormal{1}},
 Spencer Whitehead\textsuperscript{\textnormal{1}},
 Di Lu\textsuperscript{\textnormal{3}}, \\
 \textbf{Heng Ji}\textsuperscript{\textnormal{1}}, 
 \textbf{Shih-Fu Chang}\textsuperscript{\textnormal{2}} \\
  \textsuperscript{1}University of Illinois at Urbana-Champaign, 
  \textsuperscript{2}Columbia University 
   \textsuperscript{3}Dataminr \\
  \texttt{\fontfamily{pcr}\selectfont\{manling2,hengji\}@illinois.edu}, 
  \texttt{\fontfamily{pcr}\selectfont\{az2407,sc250\}@columbia.edu} \\
\url{http://blender.cs.illinois.edu/software/m2e2}
  }

\begin{document}

\maketitle

\begin{abstract}

We introduce a new task, \textbf{M}ulti\textbf{M}edia \textbf{E}vent \textbf{E}xtraction (\mmee), which aims to extract events and their arguments from multimedia documents.
We develop the first benchmark and collect a dataset of 245 multimedia news articles with extensively annotated events and arguments.\footnote{Our data and code are available at \url{http://blender.cs.illinois.edu/software/m2e2}} 
We propose a novel method, \wase~(\textbf{WASE}), that encodes structured representations of semantic information from textual and visual data into a common embedding space.
The structures are aligned across modalities by employing a weakly supervised training strategy, which enables exploiting available resources without explicit cross-media annotation.
Compared to uni-modal state-of-the-art methods, our approach achieves 4.0\% and 9.8\% absolute F-score gains on text event argument role labeling and visual event extraction.
Compared to state-of-the-art multimedia unstructured representations, we achieve 8.3\% and 5.0\% absolute F-score gains on multimedia event extraction and argument role labeling, respectively.
By utilizing images, we
extract 21.4\% more event mentions than traditional text-only methods.

\end{abstract}

\section{Introduction \label{sec:intro}}

\begin{figure}[htbp]
\centering
\centering
\includegraphics[width=1.0\linewidth]{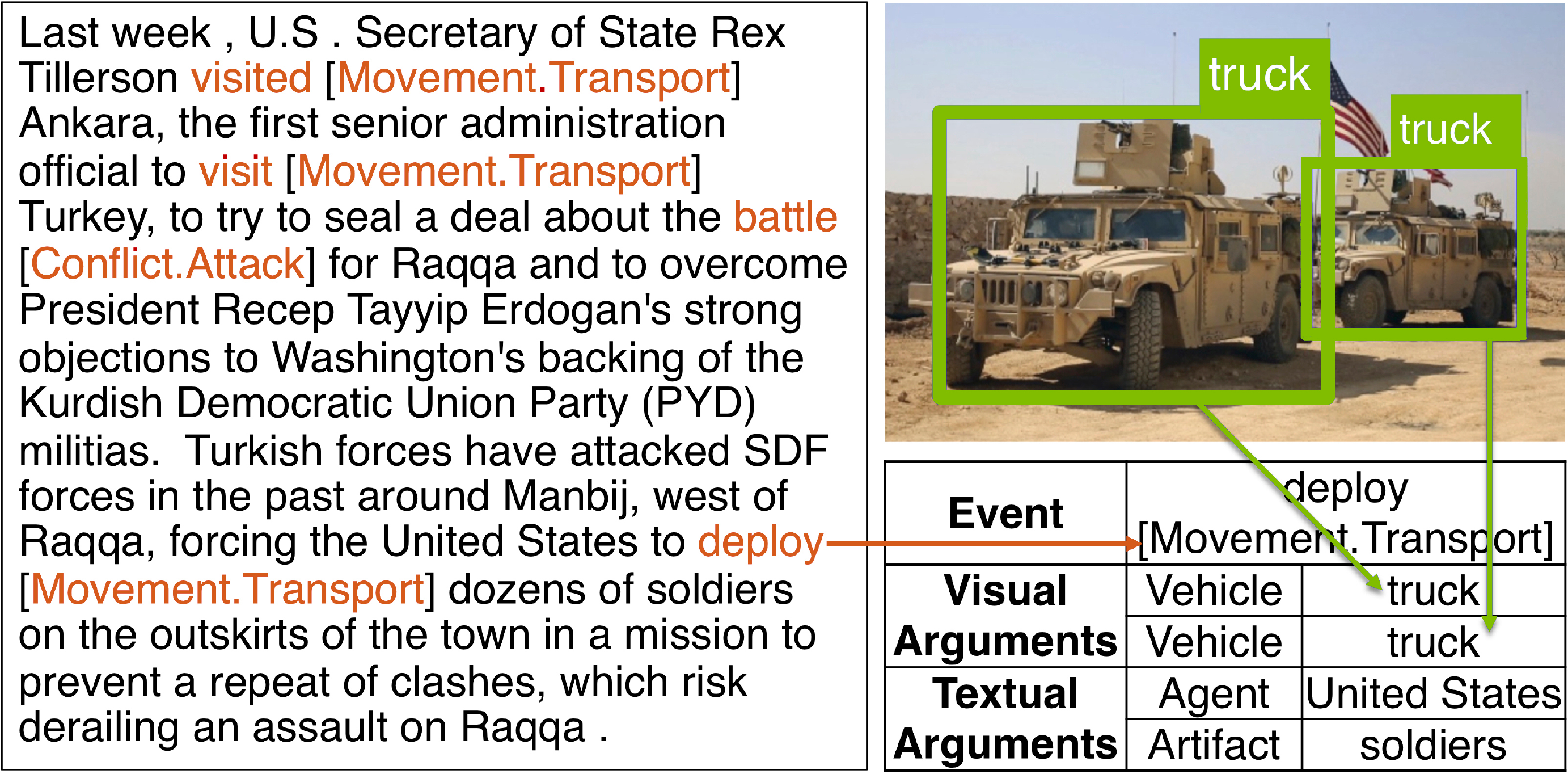}
\caption{An example of Multimedia Event Extraction. An event mention and some event arguments (\emph{Agent} and \emph{Person}) are extracted from text, while the vehicle arguments can only be extracted from the image.
}
\label{img:task_def2}
\end{figure}

Traditional event extraction methods target a single modality, such as text~\cite{wadden2019entity}, images~\cite{yatskar2016situation} or videos~\cite{ye2015eventnet, caba2015activitynet, soomro2012ucf101}. However, the practice of contemporary journalism~\citep{Mitchell1998}  distributes news via multimedia. 
By randomly sampling 100 multimedia news articles from the Voice of America (VOA), we find that 33\% of images in the articles contain visual objects that serve as event arguments and are not mentioned in the text.
Take \figurename~\ref{img:task_def2} as an example, we can extract the  \emph{Agent} and \emph{Person} arguments of the \emph{Movement.Transport} event from text, but can extract the \emph{Vehicle} argument only from 
the image.
Nevertheless, event extraction is independently studied in Computer Vision (CV) and Natural Language Processing (NLP), with major differences in task definition, data domain, methodology, and terminology. 
Motivated by the complementary and holistic nature of multimedia data, we propose \textbf{M}ulti\textbf{M}edia \textbf{E}vent \textbf{E}xtraction (\textbf{\mmee}), a new task that aims to jointly extract events and arguments from multiple modalities.
We construct the first benchmark and evaluation dataset for this task, which consists of 245 fully annotated news articles.

We propose the first method, \wase~(\textbf{WASE}), for extracting events and arguments from multiple modalities. 
Complex event structures have not been covered by existing multimedia representation methods~\cite{Wu2019UniVSERV, faghri2018vse++, karpathy2015deep}, so we propose to learn a \emph{structured} multimedia embedding space.
More specifically, given a multimedia document, we represent each image or sentence as a graph, where each node represents an event or entity and each edge represents an argument role. The node and edge embeddings are represented in a multimedia common semantic space, as they are trained to resolve event co-reference across modalities and to match images with relevant sentences. This enables us to jointly classify events and argument roles from both modalities.
A major challenge is the lack of multimedia event argument annotations, which are costly to obtain due to the annotation complexity. Therefore, we propose a weakly supervised framework, which takes advantage of annotated uni-modal corpora to separately learn visual and textual event extraction, and uses an image-caption dataset to align the modalities.

We evaluate WASE on the new task of \mmee. 
Compared to the state-of-the-art uni-modal methods and multimedia flat representations, our method 
significantly outperforms on both event extraction and argument role labeling tasks in all settings. 
Moreover, it extracts 21.4\% more event mentions than text-only baselines. 
The training and evaluation are done on heterogeneous data sets from multiple sources, domains and data modalities, demonstrating the scalability and transferability of the proposed model.
In summary, this paper makes the following contributions: 
\begin{itemize}
\item We propose a new task, MultiMedia Event Extraction, and construct the first annotated news dataset as a benchmark to support deep analysis of cross-media events.

\item We develop a weakly supervised training framework, which utilizes existing single-modal annotated corpora, and enables joint inference without cross-modal annotation. 

\item Our proposed method, WASE, is the first to leverage structured representations and graph-based neural networks for multimedia common space embedding.

\end{itemize}

\section{Task Definition \label{sec:task&dataset}}
\subsection{Problem Formulation \label{sec:task}}

Each input document consists of a set of images $\mathcal{M} = \{m_1,m_2,\dots\}$ and a set of sentences $\mathcal{S} = \{s_1,s_2,\dots\}$. 
Each sentence $s$ can be represented as a sequence of tokens $s = (w_1,w_2,\dots)$, where $w_i$ is a token from the document vocabulary $\mathcal{W}$.
The input also includes a set of entities $\mathcal{T}=\{t_1,t_2,\dots\}$ extracted from the document text. An entity is an individually unique object in the real world, such as a person, an organization, a facility, a location, a geopolitical entity, a weapon, or a vehicle.
The objective of 
\mmee is twofold: 

\textbf{Event Extraction}: Given a multimedia document, extract a set of event mentions, where each event mention $e$ has a type $y_e$ and is grounded on a text trigger word $w$ or an image $m$ or both, i.e.,
\begin{equation*}
\begin{aligned}
e = (y_e, \{w, m\}).
\end{aligned}
\end{equation*}
Note that for an event, $w$ and $m$ can both exist, which means the visual event mention and the textual event mention refer to the same event. 
For example in \figurename~\ref{img:task_def2}, 
\textit{deploy} indicates the same \textit{Movement.Transport} event as the image.
We consider the event $e$ as \textbf{text-only} event if it only has textual mention $w$, and as \textbf{image-only} event if it only contains visual mention $m$, and as \textbf{multimedia} event if both $w$ and $m$ exist. 

\textbf{Argument Extraction}: The second task is to extract a set of arguments of event mention $e$. Each argument $a$ has an argument role type $y_{a}$, and  is grounded on a text entity $t$ or an image object  $o$ (represented as a bounding box), or both, 
\begin{align*}
    a = \left(y_{a}, \{t, o\}\right).
\end{align*}
{The arguments of visual and textual event mentions are merged if they refer to the same real-world event, as shown in \figurename~\ref{img:task_def2}. 
}

\subsection{The \textbf{M}\textsuperscript{2}\textbf{E}\textsuperscript{2} Dataset \label{sec:dataset}}

We define multimedia newsworthy event types by exhaustively mapping between the event ontology in NLP community for the news domain (ACE\footnote{\url{https://catalog.ldc.upenn.edu/ldc2006T06}}) and the event ontology in CV community for general domain (imSitu~\cite{yatskar2016situation}). They cover the largest event training resources in each community.
\tablename~\ref{table:types} shows the selected complete intersection, which contains 8 ACE types (i.e., 24\% of all ACE types), mapped to 98 imSitu types (i.e., 20\% of all imSitu types). We expand the ACE event role set by adding visual arguments from imSitu, such as \emph{instrument}, bolded in Table~\ref{table:types}. This set encompasses 52\% ACE events in a news corpus, which indicates that the selected eight types are salient in the news domain. 
We reuse these existing ontologies because they enable us to train event and argument classifiers for both modalities without requiring joint multimedia event annotation as training data.

\begin{table}[!hbt]
\small
\centering
\setlength\tabcolsep{2pt}
\setlength\extrarowheight{2pt}
\begin{tabularx}{1\linewidth}{| p{0.38\linewidth} | p{0.58\linewidth} |}
\hline
\textbf{Event Type} & \textbf{Argument Role}  \\
\hline
{Movement.Transport (223$|$53)} & Agent (46$|$64), Artifact (179$|$103), Vehicle (24$|$51), Destination (120$|$0), Origin (66$|$0)  \\  
\hline
{Conflict.Attack (326$|$27)} & Attacker (192$|$12), Target (207$|$19),  Instrument (37$|$15), Place (121$|$0)  \\
\hline
{Conflict.Demonstrate (151$|$69)} & Entity (102$|$184), \textbf{Police} (3$|$26), \textbf{Instrument} (0$|$118), Place (86$|$25)  \\
\hline
{Justice.ArrestJail (160$|$56)} & Agent (64$|$119), Person (147$|$99), \textbf{Instrument} (0$|$11), Place (43$|$0)  \\
\hline
{Contact.PhoneWrite (33$|$37)} & Entity (33$|$46), \textbf{Instrument} (0$|$43), Place (8$|$0)  \\
\hline
{Contact.Meet (127$|$79)} & Participant (119$|$321), Place (68$|$0)  \\
\hline
{Life.Die \quad\quad\quad {(244$|$64)}} & Agent {(39$|$0)}, Instrument {(4$|$2)}, Victim  {(165$|$155)}, Place {(54$|$0)}  \\
\hline
{Transaction. \quad\quad TransferMoney (33$|$6)} & Giver~(19$|$3), Recipient (19$|$5), \quad\quad\quad\quad Money (0$|$8)   \\
\hline
\end{tabularx}
\caption{Event types and argument roles in \mmee, with expanded ones in bold. 
Numbers in parentheses represent the counts of textual and visual events/arguments.
}
\label{table:types}
\end{table}

We collect 108,693 multimedia news articles from the Voice of America (VOA) website~\footnote{\url{https://www.voanews.com/}} 2006-2017, covering a wide range of newsworthy topics such as military, economy and health. We select 245 documents as the annotation set based on three criteria: 
(1) Informativeness: articles with more event mentions;
(2) Illustration: articles with more images ($>4$);
(3) Diversity: articles that balance the event type distribution regardless of true frequency.
The data statistics are shown in \tablename~\ref{table:dataset_statistics}. Among all of these events, 192 textual event mentions and 203 visual event mentions can be aligned as 309 cross-media event mention pairs.
The dataset can be divided into 1,105 text-only event mentions, 188 image-only event mentions, and 395 multimedia event mentions.

\begin{table}[!htb]
\centering
\small
\setlength\tabcolsep{3pt}
\setlength\extrarowheight{2pt}
\begin{tabularx}{1\linewidth}{|p{0.18\linewidth}|p{0.13\linewidth}|p{0.13\linewidth}|p{0.13\linewidth}|p{0.13\linewidth}|p{0.13\linewidth}|}
\hline
\multicolumn{2}{|c|}{\textbf{Source}} & \multicolumn{2}{c|}{\textbf{Event Mention}} & \multicolumn{2}{c|}{\textbf{Argument Role}} \\
\cline{1-6}
 sentence  & image & textual & visual & textual & visual  
\\ 
\hline
6,167 & 1,014 & 1,297 & 391 & 1,965 & 1,429 \\
\hline
\end{tabularx}

\caption{\mmee~data statistics. 
}
\label{table:dataset_statistics}
\vspace{-9pt}
\end{table}

We follow the ACE event annotation guidelines~\cite{walker2006ace} for textual event and argument annotation, and design an annotation guideline~\footnote{\url{http://blender.cs.illinois.edu/software/m2e2/ACL2020_M2E2_annotation.pdf}} for multimedia events annotation. 

\begin{figure}[htbp]
\centering
\includegraphics[width=1.0\linewidth]{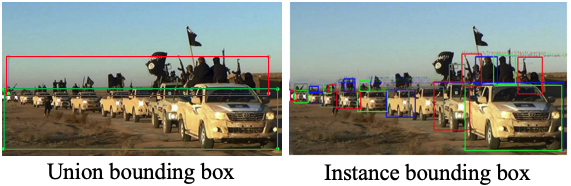}
\caption{Example of bounding boxes.}
\label{img:anno_bbox}
\end{figure}

One unique challenge in multimedia event annotation is to localize visual arguments in complex scenarios, where images include a crowd of people or a group of object. It is hard to delineate each of them using a bounding box. To solve this problem, we define two types of bounding boxes: (1) \emph{union bounding box}: for each role, we annotate the smallest bounding box covering all constituents; and (2) \emph{instance bounding box}: for each role, we annotate a set of bounding boxes, where each box is the smallest region that covers an individual participant (e.g., one person in the crowd), following the VOC2011 Annotation Guidelines\footnote{\url{http://host.robots.ox.ac.uk/pascal/VOC/voc2011/guidelines.html}}. 
\figurename~\ref{img:anno_bbox} shows an example.
Eight NLP and CV researchers complete the annotation work with two independent passes and reach an Inter-Annotator Agreement (IAA) of 81.2\%. Two expert annotators perform adjudication.

\begin{figure*}[th]
\begin{center}
\includegraphics[width=1.0\linewidth]{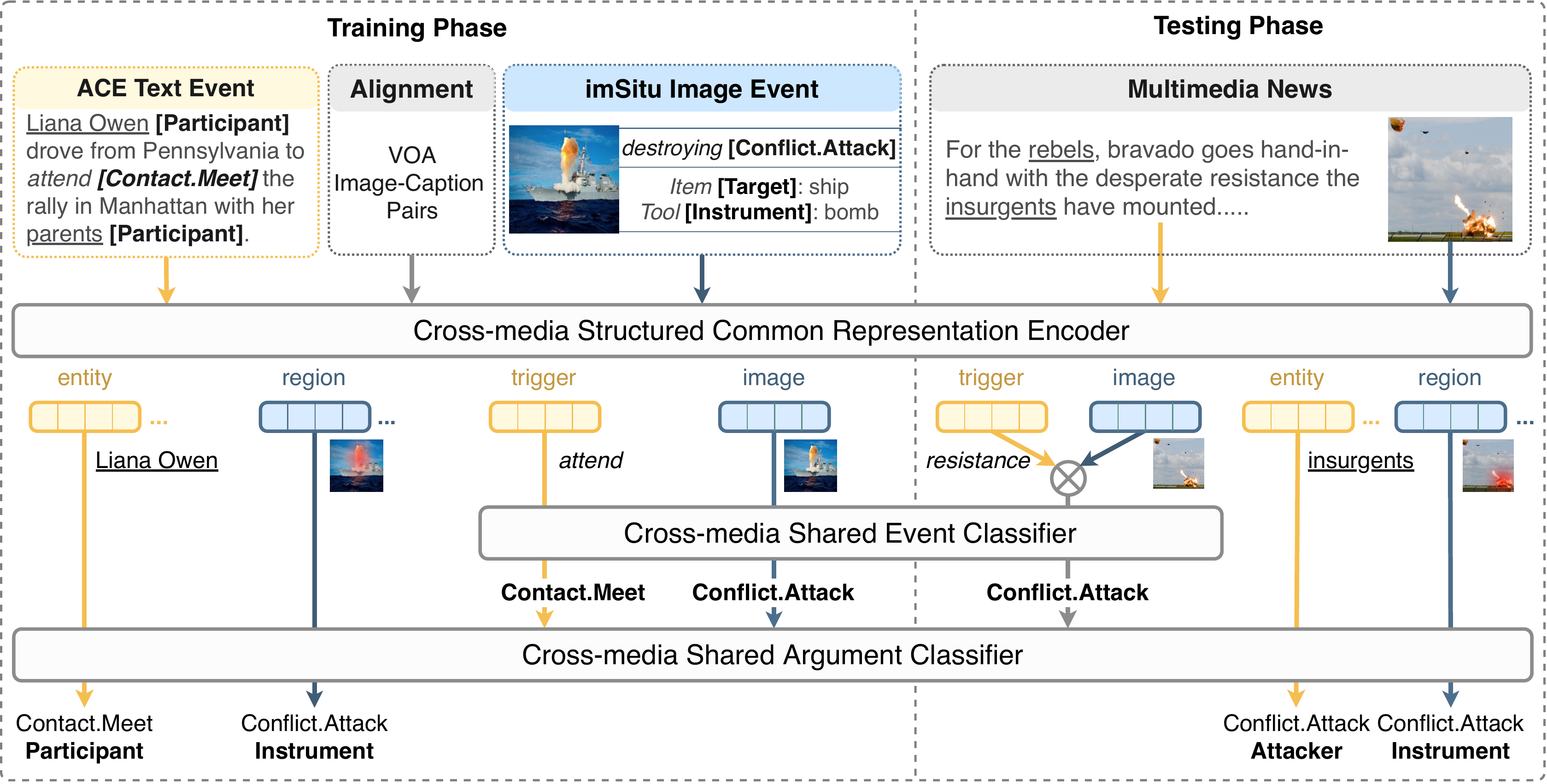}
\end{center}
\caption{Approach overview. 
During training (left), we jointly train three tasks to establish a cross-media structured embedding space. During test (right), we jointly extract events and arguments from multimedia articles. }

\label{fig:training_testing}
\end{figure*}

\begin{figure*}[th]
\begin{center}
\includegraphics[width=1\linewidth]{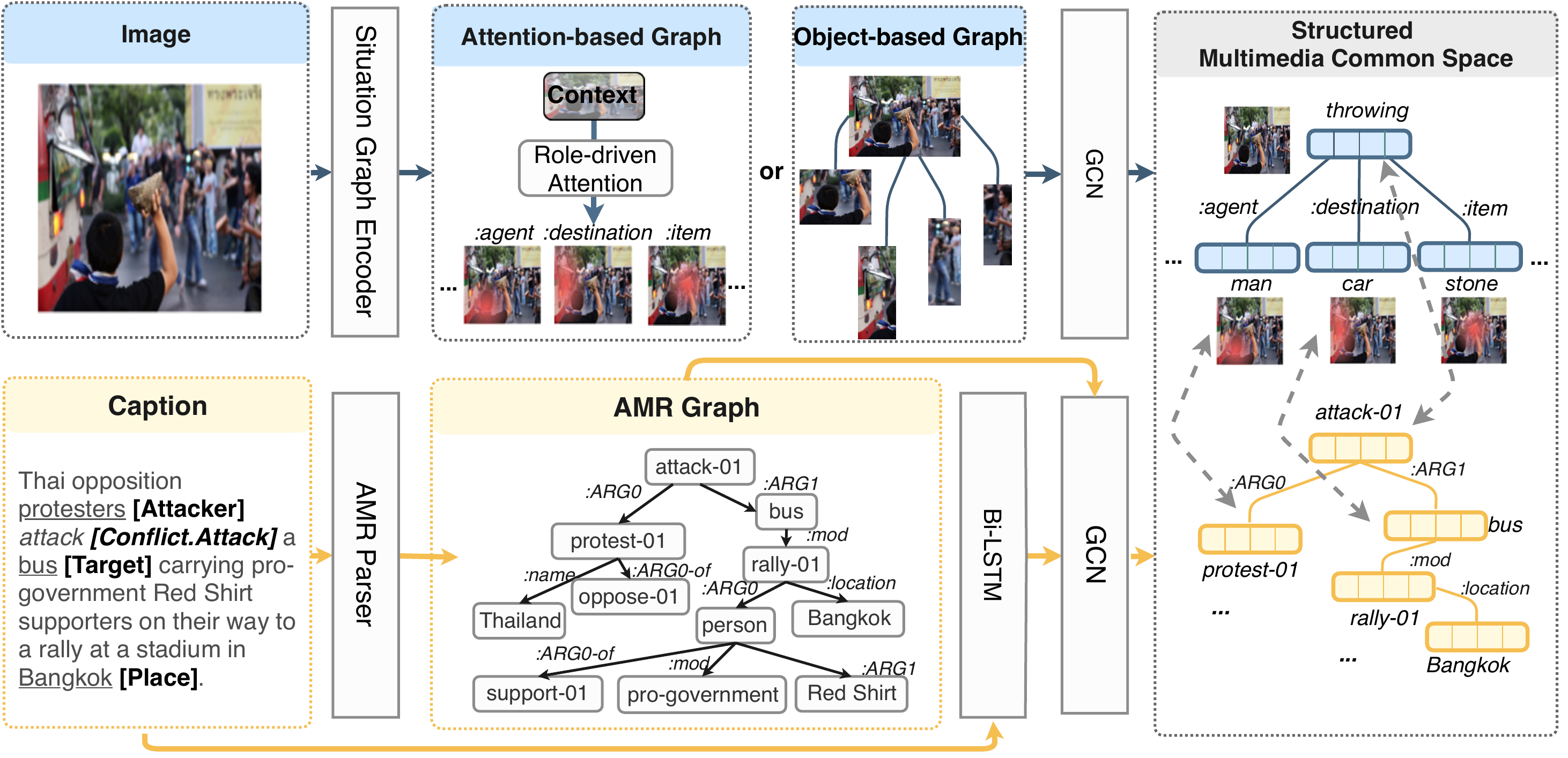}
\end{center}
\caption{Multimedia structured common space construction. Red pixels stands for attention heatmap. 
} 
\label{fig:commonspace}
\end{figure*}

\section{Method \label{sec:method}}

\subsection{Approach Overview}

As shown in \figurename~\ref{fig:training_testing}, the training phase contains three tasks: text event extraction (Section~\ref{sec:text_event}), visual situation recognition (Section~\ref{sec:image_event}), and cross-media alignment (Section~\ref{sec:crossmedia}). We learn a cross-media shared encoder, a shared event classifier, and a shared argument classifier. 
In the testing phase (Section~\ref{sec:testing}), given a multimedia news article, we encode the sentences and images into the structured common space, and jointly extract textual and visual events and arguments, followed by cross-modal coreference resolution.
 
\subsection{Text Event Extraction}\label{sec:text_event}

\textbf{Text Structured Representation:}
As shown in \figurename~\ref{fig:commonspace}, we choose Abstract Meaning Representation (AMR)~\cite{banarescu2013abstract} to represent text because it includes a rich set of 150 fine-grained semantic roles.
To encode each text sentence, we run the CAMR parser~\citep{wang-xue-pradhan:2015:NAACL-HLT, wang-xue-pradhan:2015:ACL-IJCNLP, wang-EtAl:2016:SemEval} to generate an AMR graph, based on the named entity recognition and part-of-speech (POS) tagging results from Stanford CoreNLP~\citep{manning-EtAl:2014:P14-5}.
To represent each word $w$ in a sentence $s$,  
we concatenate its pre-trained GloVe word embedding~\citep{pennington2014glove},
POS embedding, entity type embedding and position embedding. We then input the word sequence to a bi-directional long short term memory (Bi-LSTM)~\cite{graves2013speech} network to encode the word order and get the representation of each word $\bs{w}$. Given the AMR graph, we apply a Graph Convolutional Network (GCN) \citep{kipf2016semi} to encode the graph contextual information following~\citep{liu2018jointly}:
\begin{equation}
    \begin{aligned}
        \bs{w}^{(k+1)}_{i} = f(\sum_{j \in \mathcal{N}(i)} g_{ij}^{(k)} (\bs{W}_{E(i,j)} \bs{w}^{(k)}_{j} + \bs{b}^{(k)}_{E(i,j)})),
    \end{aligned}
\label{eq:graph_text}
\end{equation}
where $\mathcal{N}(i)$ is the neighbour nodes of $w_i$ in the AMR graph, $E(i,j)$ is the edge type between $w_i$ and $w_j$,  $g_{ij}$ is the gate following~\citep{liu2018jointly}, $k$ represents GCN layer number, and $f$ is the Sigmoid function.
$\bs{W}$ and $\bs{b}$ denote parameters of neural layers in this paper.
We take the hidden states of the last GCN layer for each word as the common-space representation $\bs{w}^{\mathbb{C}}$, where $\mathbb{C}$ stands for the common (multimedia) embedding space.
For each entity $t$, we obtain its representation $\bs{t}^{\mathbb{C}}$ by averaging the embeddings of its tokens.

\noindent\textbf{Event and Argument Classifier:}
We classify each word $w$ into event types $y_e$\footnote{We use BIO tag schema to decide trigger word boundary, i.e., adding prefix \textit{B-} to the type label to mark the beginning of a trigger, \textit{I-} for inside, and \textit{O} for none. } 
and classify each entity $t$ into argument role $y_a$:
\begin{equation}
\begin{aligned}
    P(y_e|w) &= \frac{\exp{ \left(\bs{W}_{e}\bs{w}^{\mathbb{C}}+\bs{b}_{e}\right) }}{\sum_{e'} \exp{ \left(\bs{W}_{e'}\bs{w}^{\mathbb{C}}+\bs{b}_{e'}\right)}}, \\
    P(y_a|t) &= \frac{\exp (\bs{W}_{a}[\bs{t}^{\mathbb{C}};\bs{w}^{\mathbb{C}}]+\bs{b}_{a}) }{\sum_{a'} \exp (\bs{W}_{a'}[\bs{t}^{\mathbb{C}};\bs{w}^{\mathbb{C}}]+\bs{b}_{a'})} . \\
\label{eq:text_classifier}
\end{aligned}
\end{equation}
We take ground truth text entity mentions as input following~\cite{ji2008refining} during training, and obtain testing entity mentions using a named entity extractor~\citep{LinACL2019}.

\subsection{Image Event Extraction}\label{sec:image_event}

\textbf{Image Structured Representation: }\label{sec:graph_image}
To obtain image structures similar to AMR graphs, and inspired by \textit{situation recognition}~\cite{yatskar2016situation}, we represent each image with a \emph{situation graph}, that is a star-shaped graph as shown in Figure~\ref{fig:commonspace}, 
where the central node is labeled as a verb $v$ (e.g., \textit{destroying}), 
and the neighbor nodes are arguments labeled as $\{(n, r)\}$, where $n$ is a noun (e.g., \textit{ship}) derived from WordNet synsets~\cite{miller1995wordnet} to indicate the entity type, and $r$ indicates the role (e.g., \textit{item}) played by the entity in the event, based on FrameNet~\cite{fillmore2003background}. We develop two methods to construct situation graphs from images and train them using the imSitu dataset~\cite{yatskar2016situation} as follows.

\noindent\textbf{(1) Object-based Graph: }
Similar to extracting entities to get candidate arguments, we employ the most similar task in CV, object detection, and 
obtain the object bounding boxes detected by a 
Faster R-CNN
~\cite{ren2015faster} model trained on Open Images~\cite{kuznetsova2018open} with 600 object types (classes).
We employ a 
VGG-16 CNN \citep{simonyan2014very} to extract visual features of an image $\bs{m}$ and and another VGG-16 to encode the bounding boxes $\{\bs{o}_i\}$.
Then we apply a Multi-Layer Perceptron (MLP) to predict a verb embedding from $\bs{m}$ and another MLP to predict a noun embedding for each $\bs{o}_i$.
\begin{equation*}
\begin{aligned}
\bs{\hat{m}} &= {\rm MLP_{m}} (\bs{m})\;, \; \hat{\bs{o}_i} = {\rm MLP_{o}} (\bs{o}_i) 
.\end{aligned}
\end{equation*}
We compare the predicted verb embedding to all verbs $v$ in the imSitu taxonomy in order to classify the verb, and similarly compare each predicted noun embedding to all imSitu nouns $n$ which results in probability distributions:  
\vspace{-6pt}
\begin{equation*}
\begin{aligned}
P(v|m) &= \frac{\exp{ (\hat{\bs{m}}\bs{v})}}{\sum_{v'}\exp{ (\hat{\bs{m}}\bs{v}') }}, \\
P(n|o_i) &= \frac{\exp (\hat{\bs{o}_i}\bs{n})}{\sum_{n'} \exp (\hat{\bs{o}_i}\bs{n}')} ,
\end{aligned}
\end{equation*}
where $\bs{v}$ and $\bs{n}$ are word embeddings initialized with GloVE~\cite{pennington2014glove}.
We use another MLP with one hidden layer followed by Softmax ($\sigma$) to classify role $r_i$ for each object $o_i$: 
\begin{equation*}
\begin{aligned}
P(r_i|o_i) &= \sigma\big({\rm MLP_{r}} (\hat{\bs{o}_i})\big)
.\end{aligned}
\end{equation*}
Given verb $v^*$ and role-noun $(r_i^*, n_i^*)$ annotations for an image (from the imSitu corpus), we define the situation loss functions: 
\begin{equation*} \label{eq:loss_img_obj}
\begin{aligned}
& \mathcal{L}_{v} = - \log P(v^*|m), \\
& \mathcal{L}_{r} = - \log (P(r_i^*|o_i) + P(n_i^*|o_i)).
\end{aligned}
\end{equation*}

\noindent\textbf{(2) Attention-based Graph: }
State-of-the-art object detection methods only cover a limited set of object types, such as 600 types defined in  Open Images. 
Many salient objects such as \textit{bomb}, \textit{stone} and \textit{stretcher} are not covered in these ontologies. 
Hence, we propose an open-vocabulary alternative to the object-based graph construction model. To this end, we construct a role-driven attention graph, where each argument node is derived by a spatially distributed attention (heatmap) conditioned on a role $r$. 
More specifically, we use a VGG-16 CNN to extract a $7 \times 7$ convolutional feature map 
for each image $m$, which can be regarded as 
attention \emph{keys} $\bs{k}_{i}$ for $7 \times 7$ local regions.
Next, for each role $r$ defined in the situation recognition ontology 
(e.g., \textit{agent}),
we build an attention \textit{query} vector $\bs{q}_{r}$ by concatenating role embedding $\bs{r}$ with the image feature $\bs{m}$ as context and apply a fully connected layer: 
\begin{equation*}
    \begin{aligned}
        \bs{q}_{r} = \bs{W}_q[\bs{r};\bs{m}] + \bs{b}_q.
    \end{aligned}
\end{equation*}
Then, we compute the dot product of each query with all keys, followed by Softmax, which forms a heatmap $\bs{h}$ on the image, i.e.,
\begin{equation*}
    \begin{aligned}
        h_i = \frac{\exp (\bs{q}_r\bs{k}_i)}{\sum_{j \in {7 \times 7} } \exp (\bs{q}_r\bs{k}_j)}.  
    \end{aligned}
\end{equation*}
We use the heatmap to obtain a weighted average of the feature map to represent the argument $o_r$ of each role $r$ in the visual space: 
\begin{equation*}
\begin{aligned}
\bs{o}_r = \sum_{i} h_i \bs{m}_i.
\end{aligned}
\end{equation*}
Similar to the object-based model, we embed $\bs{o}_r$ to $\hat{\bs{o}_r}$, compare it to the imSitu noun embeddings to define a distribution, and define a classification loss function. The verb embedding $\hat{\bs{m}}$ and the verb prediction probability $P(v|m)$ and loss are defined in the same way as in the object-based method.

\noindent\textbf{Event and Argument Classifier: }
We use either the object-based or attention-based formulation and pre-train it on the imSitu dataset~\cite{yatskar2016situation}. Then we apply a GCN to obtain the structured embedding of each node in the common space, similar to Equation 1. 
This yields $\bs{m}^{\mathbb{C}}$ and $\bs{o}_i^{\mathbb{C}}$.
We use the same classifiers as defined in Equation 2 
to classify each visual event and argument using the common space embedding:
\begin{equation} \label{eq:visual_classifiers}
\begin{aligned}
    P(y_e|m) &= \frac{\exp (\bs{W}_{e}\bs{m^{\mathbb{C}}}+\bs{b}_{e})}{\sum_{e'} \exp (\bs{W}_{e'}\bs{m^{\mathbb{C}}}+\bs{b}_{e'})} , \\
    P(y_a|o) &= \frac{\exp (\bs{W}_{a}[\bs{o^{\mathbb{C}}};\bs{m^{\mathbb{C}}}]+\bs{b}_{a})}{\sum_{a'} \exp  (\bs{W}_{a'}[\bs{o^{\mathbb{C}}};\bs{m^{\mathbb{C}}}]+\bs{b}_{a'})} . \\
\end{aligned}
\end{equation}
 
\subsection{Cross-Media Joint Training} \label{sec:crossmedia}

In order to make the event and argument classifier shared across modalities, the image and text graph should be encoded to the same space. 
However, it is extremely costly to obtain the parallel text and image event annotation. Hence, we use event and argument annotations in separate modalities (i.e., ACE and imSitu datasets) to train classifiers, and simultaneously use VOA news image and caption pairs to align the two modalities. To this end, we learn to embed the nodes of each image graph close to the nodes of the corresponding caption graph, and far from those in irrelevant caption graphs. Since there is no ground truth alignment between the image nodes and caption nodes, we 
use image and caption pairs for weakly supervised training, to learn a soft alignment from each words to image objects and vice versa. 
\begin{equation*}
\begin{aligned}
\alpha_{ij} = \frac{\exp{(\bs{w^{\mathbb{C}}_i}\bs{o^{\mathbb{C}}_j})}}{\sum_{j'} \exp{(\bs{w^{\mathbb{C}}_i}\bs{o^{\mathbb{C}}_{j'}})}}, 
\beta_{ji} = \frac{\exp{(\bs{w^{\mathbb{C}}_i}\bs{o^{\mathbb{C}}_j})}}{\sum_{i'} \exp{(\bs{w^{\mathbb{C}}_{i'}}\bs{o^{\mathbb{C}}_{j}})}} ,
\end{aligned}
\end{equation*}
where $w_i$ indicates the $i^{th}$ word in caption sentence $s$ and $o_j$ represents the $j^{th}$ object of image $m$. 
Then, we compute a weighted average of softly aligned nodes for each node in other modality, i.e.,
\begin{equation} \label{eq:aligned_features}
\begin{aligned}
\bs{w}'_i = \sum_{j} \alpha_{ij} \bs{o}^{\mathbb{C}}_j \;,\; 
\bs{o}'_j = \sum_{i} \beta_{ji}  \bs{w}^{\mathbb{C}}_i .
\end{aligned}
\end{equation}
We define the alignment cost of the image-caption pair as the Euclidean distance between each node to its aligned representation, 
\begin{equation*}
    \begin{aligned}
        \langle s,m \rangle &= \sum_{i} ||\bs{w}_i-\bs{w}'_i||_2^2 + \sum_{j} ||\bs{o}_j-\bs{o}'_j||_2^2 \\
    \end{aligned}
\end{equation*}
We use a triplet loss to pull relevant image-caption pairs close while pushing irrelevant ones apart:
\begin{equation*} 
    \begin{aligned}
        \mathcal{L}_{c} &= \max (0, 1+\langle s,m \rangle-\langle s,m^- \rangle), 
    \end{aligned}
\end{equation*}
where $m^-$ is a randomly sampled negative image that does not match $s$. Note that in order to learn the alignment between the image and the trigger word, we treat the image as a special object when learning cross-media alignment. 

The common space enables the event and argument classifiers to share weights across modalities, and be trained jointly on the ACE and imSitu datasets, by minimizing the following objective functions: 
\begin{equation*}
\begin{aligned}
    & \mathcal{L}_{e} = - \sum_{w} \log P(y_e|w) - \sum_{m}  \log P(y_e|m), \\
    & \mathcal{L}_{a} = - \sum_{t} \log P(y_a|t)  - \sum_{o} \log P(y_a|o),
\end{aligned}
\end{equation*}
All tasks are jointly optimized: 
\begin{equation*}
    \begin{aligned}
        \mathcal{L} = \mathcal{L}_{v} + \mathcal{L}_{r} + \mathcal{L}_{e} + \mathcal{L}_{a} + \mathcal{L}_{c}
    \end{aligned}
\end{equation*}

\newcommand{\tabincell}[2]{\begin{tabular}{@{}#1@{}}#2\end{tabular}}

\begin{table*}[!htb]
\centering
\small
\setlength\tabcolsep{3pt}
\setlength\extrarowheight{2pt}

\begin{tabular}{|c|c|c c c|c c c|c c c|c c c|c c c|c c c|}
\hline
\multirow{3}{*}{\rotatebox{-90}{\hspace{-0.25cm}\textbf{Training}}} & \multirow{3}{*}{\textbf{Model}}
& 
\multicolumn{6}{c|}{\textbf{Text-Only Evaluation}} & \multicolumn{6}{c|}{\textbf{Image-Only Evaluation}} & \multicolumn{6}{c|}{\textbf{Multimedia Evaluation}}
\\ \cline{3-20}
& & \multicolumn{3}{c|}{\textbf{Event Mention}} & \multicolumn{3}{c|}{\textbf{Argument Role}} & \multicolumn{3}{c|}{\textbf{Event Mention}} & \multicolumn{3}{c|}{\textbf{Argument Role}} & \multicolumn{3}{c|}{\textbf{Event Mention}} & \multicolumn{3}{c|}{\textbf{Argument Role}}  
\\ \cline{3-20}
& & \textbf{$P$ }  & \textbf{$R$} & \textbf{$F_{1}$} & 
\textbf{$P$} & \textbf{$R$} & \textbf{$F_{1}$}  & 
\textbf{$P$} & \textbf{$R$} & \textbf{$F_{1}$}  & 
\textbf{$P$} & \textbf{$R$} & \textbf{$F_{1}$}  & 
\textbf{$P$} & \textbf{$R$} & \textbf{$F_{1}$}  & 
\textbf{$P$} & \textbf{$R$} & \textbf{$F_{1}$}  
\\ \hline
\multirow{3}{*}{~\rotatebox{-90}{\hspace{-0.25cm}\textbf{Text}}} 
& JMEE 
& 42.5 & 58.2 & 48.7 & 22.9 & 28.3 & 25.3 
& -  & - & - & - & - & -  
& 42.1 & 34.6 & 38.1 & 21.1 & 12.6 & 15.8 \\ 
& GAIL 
& {43.4} & 53.5 & 47.9 & 23.6 & 29.2 & 26.1 
& -  & - & - & - & - & -  
& {44.0} & 32.4 & 37.3 & {22.7} & 12.8 & 16.4 \\ 
& WASE\textsuperscript{$\mathbb{T}$}
& 42.3 & 58.4 & 48.2 & 21.4 & 30.1 & 24.9  
& -  & - & - & - & - & -  
& 41.2 & 33.1 & 36.7 & 20.1 & 13.0 & 15.7 \\ 
\hline
\multirow{2}{*}{\rotatebox{-90}{\hspace{-0.25cm}\textbf{Image}}} 
& WASE\textsuperscript{$\mathbb{I}$}\textsubscript{att}
&  - & - & - & - & - & - 
& 29.7 & 61.9 & 40.1 & 9.1 & 10.2 & 9.6  
& 28.3 & 23.0 & 25.4 & 2.9 & 6.1 & 3.8 \\
& WASE\textsuperscript{$\mathbb{I}$}\textsubscript{obj}
&  - & - & - & - & - & - 
& 28.6 & 59.2 & 38.7 & 13.3 & 9.8 & 11.2
& 26.1 & 22.4 & 24.1 & 4.7  & 5.0 & 4.9 \\
\hline
\multirow{5}{*}{\rotatebox{-90}{\hspace{-0.25cm}\textbf{Multimedia}}} 
& VSE-C 
& 33.5 & 47.8 & 39.4 & 16.6 & 24.7 & 19.8 
& 30.3 & 48.9 & 26.4 & 5.6  & 6.1  & 5.7 
& 33.3 & 48.2 & 39.3 & 11.1 & 14.9 & 12.8 \\ 
 & Flat\textsubscript{att}
& 34.2 & 63.2 & 44.4 & 20.1 & 27.1 & 23.1
& 27.1 & 57.3 & 36.7 & 4.3  & 8.9  & 5.8  
& 33.9 & 59.8 & 42.2 & 12.9 & 17.6 & 14.9 \\ 
& Flat\textsubscript{obj}
& 38.3 & 57.9 & 46.1 & 21.8 & 26.6 & 24.0   
& 26.4 & 55.8 & 35.8 &  9.1 & 6.5 & 7.6   
& 34.1 & 56.4 & 42.5 & 16.3 & 15.9 & 16.1 \\ 
& WASE\textsubscript{att}
& 37.6 & {66.8} & 48.1 & {27.5} & {33.2} & \textbf{30.1} 
& 32.3 & {63.4} & 42.8 & 9.7 & {11.1} & 10.3 
& 38.2 & {67.1} & 49.1 & 18.6 & {21.6} & \textbf{19.9} \\ 
& WASE\textsubscript{obj}
& {42.8} & 61.9 & \textbf{50.6} & 23.5 & 30.3 & 26.4 
& {43.1} & 59.2 & \textbf{49.9} & {14.5} & 10.1 & \textbf{11.9} 
& 43.0 & {62.1} & \textbf{50.8} & {19.5} & 18.9 & 19.2 \\ 
\hline
\end{tabular}

\caption{Event and argument extraction results (\%). We compare three categories of baselines in three evaluation settings. The main contribution of the paper is joint training and joint inference on multimedia data (bottom right).}
\label{table:result_m2e2}
\vspace{-9pt}
\end{table*}

\subsection{Cross-Media Joint Inference}\label{sec:testing}

In the test phase, our method takes a multimedia document with sentences $S=\{s_1,s_2,\dots\}$ and images $M=\{m_1, m_2,\dots,\}$ as input. We first generate the structured common embedding for each sentence and each image, and then compute pairwise similarities $\langle s, m \rangle$. We pair each sentence $s$ with the closest image $m$, and aggregate the features of each word of $s$ with the aligned representation from $m$ by weighted averaging: 
\begin{equation}
    \begin{aligned}
        \bs{w}''_i = (1 - \gamma) \bs{w}_i + \gamma \bs{w}'_i,
    \end{aligned}
\end{equation}
where $\gamma = \exp(-\langle s, m \rangle)$ and $\bs{w}'_i$ is derived from $m$ using Equation~\ref{eq:aligned_features}. We use $\bs{w}''_i$ to classify each word into an event type and to classify each entity into a role with multimedia classifiers in Equation~\ref{eq:text_classifier}. 
To this end, we define $\bs{t}''_i$ similar to $\bs{w}''_i$ but using $\bs{t}_i$ and $\bs{t}'_i$. 
Similarly, for each image $m$ we find the closest sentence $s$, compute the aggregated multimedia features $\bs{m}''$ and $\bs{o}''_i$, and feed into the shared classifiers (Equation~\ref{eq:visual_classifiers}) to predict visual event and argument roles.
Finally, we corefer the cross-media events of the same event type if the similarity $\langle s, m \rangle$ is higher than a threshold.

\section{Experiments \label{sec:experiment}}

\subsection{Evaluation Setting}

\textbf{Evaluation Metrics}
We conduct evaluation on text-only, image-only, and multimedia event mentions in M\textsuperscript{2}E\textsuperscript{2} dataset in Section~\ref{sec:dataset}.
We adopt the traditional event extraction measures, i.e., \textit{Precision}, \textit{Recall} and \textit{F}\textsubscript{1}. 
For text-only event mentions, we follow~\cite{ji2008refining,li2013joint}: a textual event mention is correct if its event type and trigger offsets match a reference trigger; and a textual event argument is correct if its event type, offsets, and role label match a reference argument. We make a similar definition for image-only event mentions: a visual event mention is correct if its event type and image match a reference visual event mention; and a visual event argument is correct if its event type, localization, and role label match a reference argument. A visual argument is correctly localized if the Intersection over Union (IoU) of the predicted bounding box with the ground truth bounding box 
is over 0.5. Finally, we define a multimedia event mention to be correct if its event type and trigger offsets (or the image) match the reference trigger (or the reference image). The arguments of multimedia events are either textual or visual arguments, and are evaluated accordingly. To generate bounding boxes for the attention-based model, we threshold the heatmap using the adaptive value of $0.75*p$, where $p$ is the peak value of the heatmap. Then we compute the tightest bounding box that encloses all of the thresholded region. Examples are shown in \figurename~\ref{img:result_bad_correctname} and \figurename~\ref{img:result_bad_similar}.  

\textbf{Baselines}
The baselines include: (1) \textbf{Text-only} models: We use the state-of-the-art model JMEE~\citep{liu2018jointly} and GAIL~\cite{Zhang2019} for comparison. We also evaluate the effectiveness of cross media joint training by including a version of our model trained only on ACE, denoted as WASE\textsuperscript{$\mathbb{T}$}.
(2) \textbf{Image-only} models: Since we are the first to extract newsworthy events, and the most similar work \textit{situation recognition} can not localize arguments in images, we use our model trained only on image corpus as baselines. 
Our visual branch has two versions, object-based and attention-based, denoted as WASE\textsuperscript{$\mathbb{I}$}\textsubscript{obj} and WASE\textsuperscript{$\mathbb{I}$}\textsubscript{att}.
(3) \textbf{Multimedia} models: 
To show the effectiveness of structured embedding, we include a baseline by removing the text and image GCNs from our model, which is denoted as Flat. The Flat baseline ignores edges and treats images and sentences as sets of vectors. We also compare to the state-of-the-art cross-media common representation model, Contrastive Visual Semantic Embedding VSE-C~\cite{shi2018learning}, by training it the same way as WASE. 

\textbf{Parameter Settings} 
The common space dimension is $300$. The dimension is $512$ for image position embedding and feature map, and $50$ for word position embedding, entity type embedding, and POS tag embedding. The layer of GCN is $3$. 

\subsection{Quantitative Performance}

As shown in \tablename~\ref{table:result_m2e2}, our complete methods (WASE\textsubscript{att} and WASE\textsubscript{obj}) outperform all baselines in the three evaluation settings in terms of F\textsubscript{1}. 
The comparison with other multimedia models demonstrates the effectiveness of our model architecture and training strategy.
The advantage of structured embedding is shown by the better performance over the flat baseline.
Our model outperforms its text-only and image-only variants on multimedia events, showing the inadequacy of single-modal information for complex news understanding. Furthermore, our model achieves better performance on text-only and image-only events, which demonstrates the effectiveness of multimedia training framework in knowledge transfer between modalities. 

WASE\textsubscript{obj} and WASE\textsubscript{att}, are both superior to the state of the art and each has its own advantages. 
WASE\textsubscript{obj} predicts more accurate bounding boxes since it is based on a Faster R-CNN pretrained on bounding box annotations, resulting in a higher argument precision. While WASE\textsubscript{att} achieves a higher argument recall as it is not limited by the predefined object classes of the Faster R-CNN.

\begin{table}[!htb]
\centering
\small
\setlength\tabcolsep{3pt}
\setlength\extrarowheight{2pt}
\begin{tabular}{|c|c|c|c|}
\hline
\textbf{Model} & \textbf{$P$ }(\%)  & \textbf{$R$} (\%) & \textbf{$F_{1}$}   (\%)
\\ 
\hline
rule\_based & 
10.1 & 100 & 18.2 \\
VSE & 
31.2 & 74.5 & 44.0 \\
Flat\textsubscript{att} & 
33.1 & 73.5 & 45.6 \\
Flat\textsubscript{obj} & 
34.3 & 76.4 & 47.3 \\
\hline
{WASE}\textsubscript{att} & 
39.5 & 73.5 & 51.5 \\
{WASE}\textsubscript{obj} & 
40.1 & 75.4 & 52.4 \\
\hline
\end{tabular}

\caption{Cross-media event coreference performance.}
\label{table:coreference}
\vspace{-9pt}
\end{table}

Furthermore, to evaluate the cross-media event coreference performance, 
we pair textual and visual event mentions in the same document, and calculate \textit{Precision}, \textit{Recall} and \textit{F}\textsubscript{1} 
to compare with ground truth event mention pairs\footnote{We do not use coreference clustering metrics because we only focus on mention-level cross-media event coreference instead of the full coreference in all documents.}.  
As shown in \tablename~\ref{table:coreference}, WASE\textsubscript{obj} outperforms all multimedia embedding models, as well as the rule-based baseline using event type matching. This demonstrates the effectiveness of our cross-media soft alignment. 
\subsection{Qualitative Analysis}

\begin{figure}[htbp]
\centering
\includegraphics[width=1.0\linewidth]{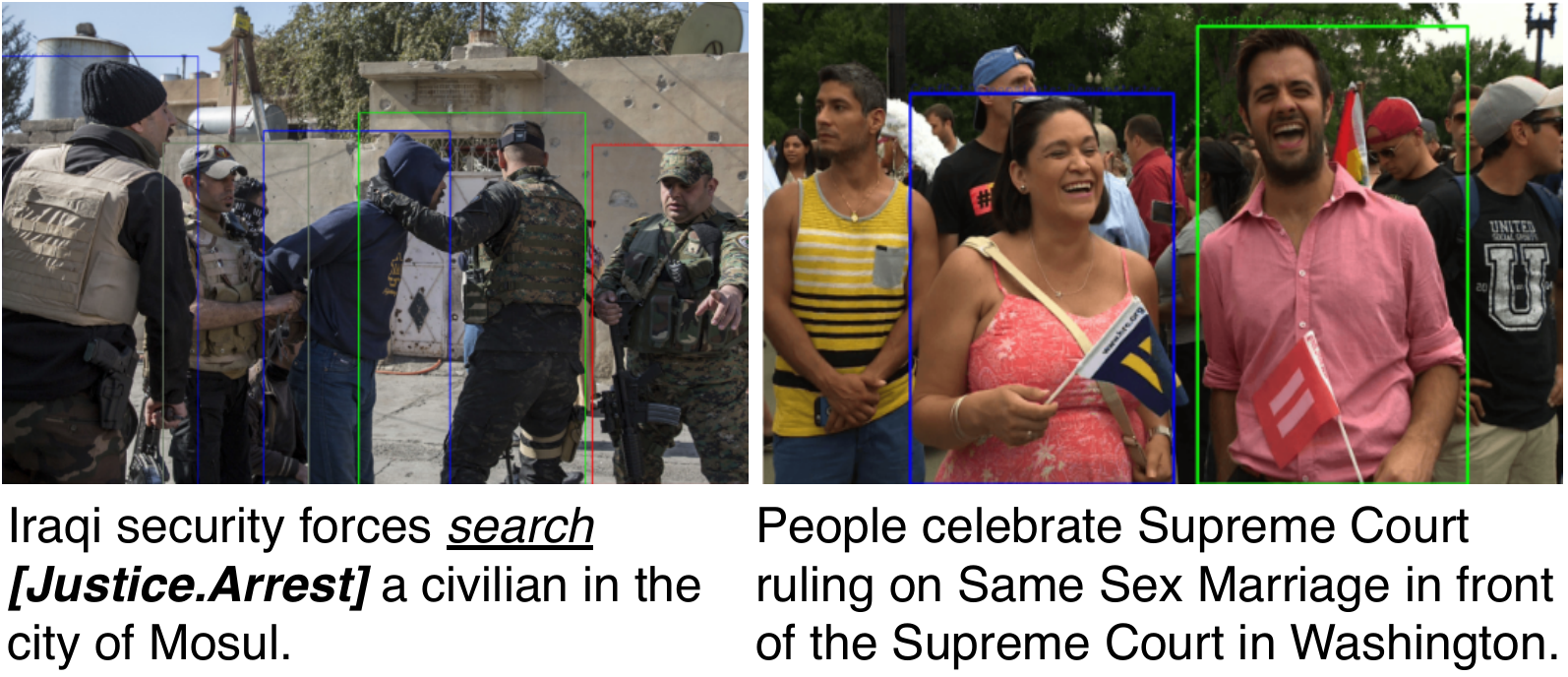}
\caption{Image helps textual event extraction, and surrounding sentence helps visual event extraction.}
\label{img:crossmedia_benefit}
\end{figure}

\begin{figure}[htbp]
\centering
\includegraphics[width=1.0\linewidth]{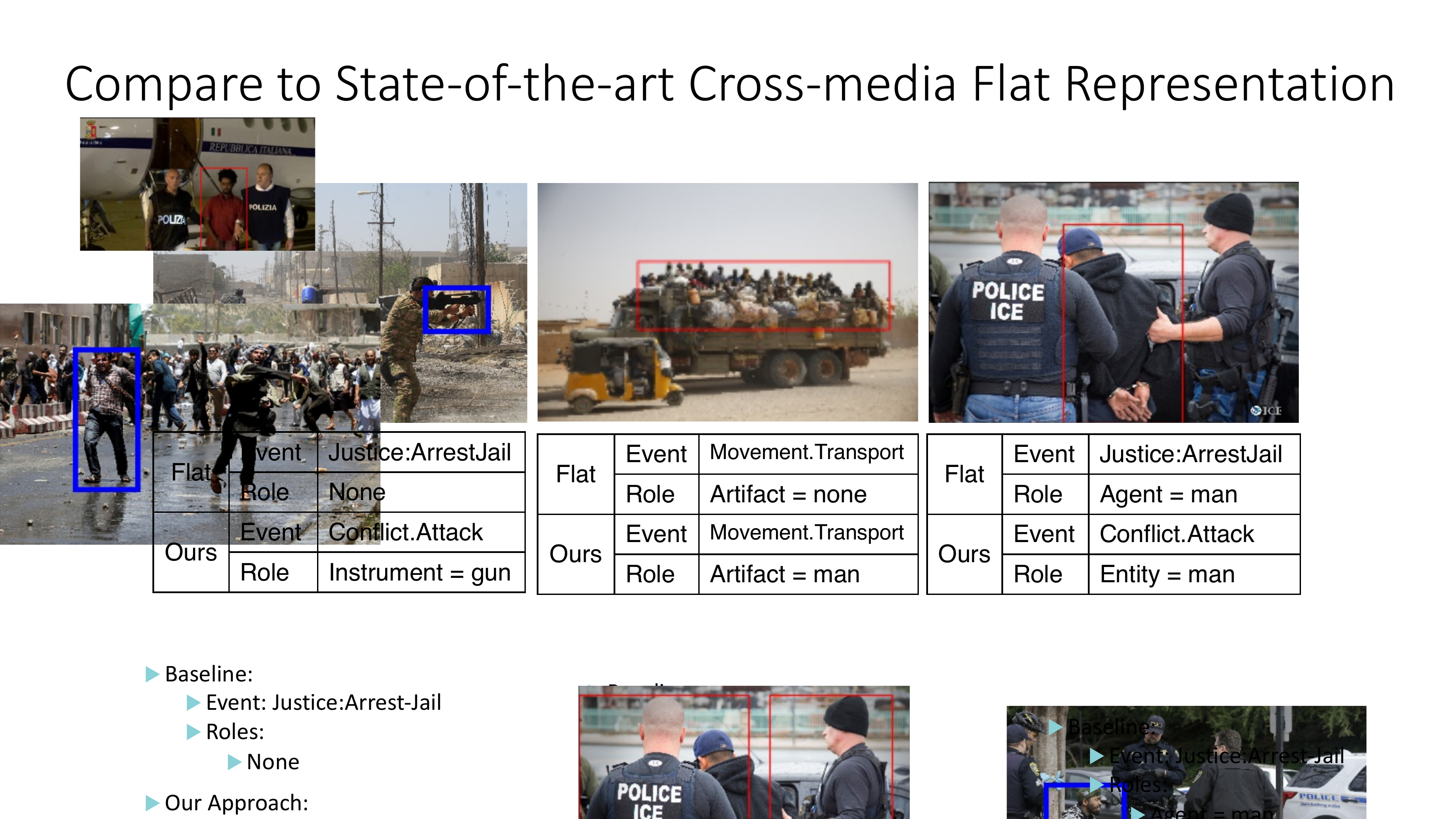}
\caption{Comparison with multimedia flat embedding. 
}
\label{img:result_compare}
\end{figure}

Our cross-media joint training approach successfully boosts both event extraction and argument role labeling performance. 
For example, in \figurename~\ref{img:crossmedia_benefit} (a), the text-only model can not extract \textit{Justice.Arrest} event, but the joint model can use the image as background to detect the event type. In \figurename~\ref{img:crossmedia_benefit} (b), the image-only model detects the image as \textit{Conflict.Demonstration}, but the sentences in the same document 
help our model not to label it as \textit{Conflict.Demonstration}. Compared with multimedia flat embedding in \figurename~\ref{img:result_compare}, WASE can learn structures such as \textit{Artifact} is on top of \textit{Vehicle}, and the person in the middle of \textit{Justice.Arrest} is \textit{Entity} instead of \textit{Agent}.

\subsection{Remaining Challenges}
\begin{figure}[htbp]
\centering
\includegraphics[width=1.0\linewidth]{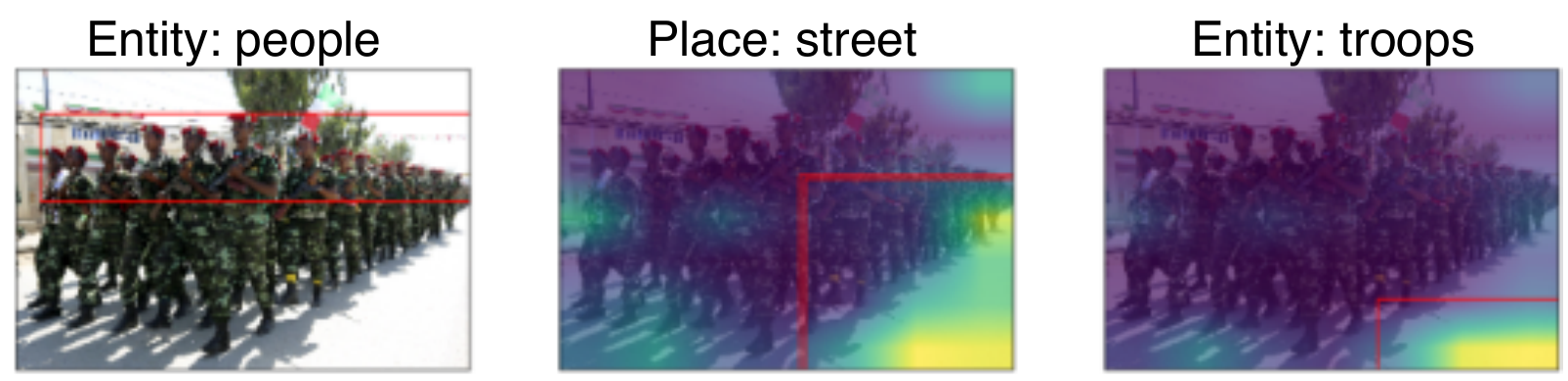}
\caption{Argument labeling error examples: correct entity name but wrong localization. 
}
\label{img:result_bad_correctname}
\end{figure}

\begin{figure}[htbp]
\centering
\includegraphics[width=1.0\linewidth]{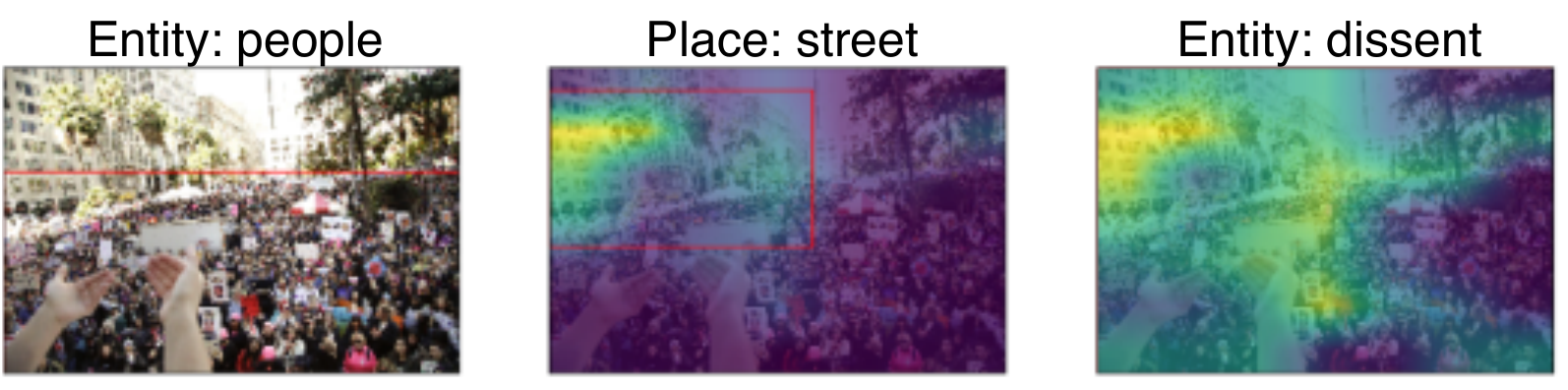}
\caption{Attention heatmaps lose focus due to large instance candidate number. 
}
\label{img:result_bad_similar}
\end{figure}

One of the biggest challenges in \mmee is localizing arguments in images. 
Object-based models suffer from the limited object types. Attention-based method is not able to precisely localize the objects for each argument, since there is no supervision on attention extraction during training. 
For example, in \figurename~\ref{img:result_bad_correctname}, the \textit{Entity} argument in the \textit{Conflict.Demonstrate} event is correctly predicted as \textit{troops}, but its localization is incorrect because \textit{Place} argument share similar attention. %
When one argument targets at too many instances,  attention heatmaps tend to lose focus and cover the whole image, as shown in \figurename~\ref{img:result_bad_similar}.


\section{Related Work \label{sec:literature}}

\textbf{Text Event Extraction} 
Text event extraction has been extensively studied for general news domain~\cite{ji2008refining,liao2011acquiring,huang2012bootstrapped,li2013joint,chen2015event, nguyen2016joint,P18-1048,D18-1156,D18-1158,Zhang2019,liu2018jointly,wang2019open,yang2019exploring,wadden2019entity}. Multimedia features has been proven to effectively improve text 
event extraction~\citep{zhang2017improving}.

\textbf{Visual Event Extraction} 
``Events" in NLP usually refer to complex events that involve multiple entities in a large span of time (e.g. protest), while in CV~\cite{chang2016bi,zhang2007semantic, ma2017joint} events are less complex single-entity activities (e.g. washing dishes) or actions (e.g. jumping). 
Visual event ontologies focus on daily life domains, such as “dogshow” and “wedding ceremony”~\citep{perera2012trecvid}. Moreover, most efforts ignore the structure of events including arguments. There are a few methods that aim to localize the agent \cite{gu2018ava, li2018recurrent, duarte2018videocapsulenet}, or classify the recipient \cite{sigurdsson2016hollywood, kato2018compositional, wu2019long} of events, but neither 
detects the complete set of arguments for an event.
The most similar to our work is Situation Recognition (SR)~\citep{yatskar2016situation, mallya2017recurrent} which predicts an event and multiple arguments from an input image, but does not localize the arguments. 
We use SR as an auxiliary task for training our visual branch, but exploit object detection and attention to enable localization of arguments. 
\citeauthor{silberer2018grounding} redefine the problem of visual argument role labeling with event types and bounding boxes as input. Different from their work, we extend the problem scope to including event identification and coreference, and further advance argument localization by proposing an attention framework which does not require bounding boxes for training nor testing.

\textbf{Multimedia Representation}
Multimedia common representation has attracted much attention recently~\cite{toselli2007viterbi,weegar2015linking,hewitt2018learning, chen2019uniter,liu2019focus,Su_2019_ICCV,Sarafianos_2019_ICCV,sun2019videobert,tan2019lxmert,li2019unicoder,li2019visualbert,lu2019vilbert,sun2019contrastive,rahman2019m,su2019vl}. However, previous methods focus on aligning images with their captions, or regions with words and entities, but ignore structure and semantic roles.
UniVSE~\cite{Wu2019UniVSERV} incorporates entity attributes and relations into cross-media alignment, but does not capture graph-level structures of images or text.

\section{Conclusions and Future Work \label{sec:conclusion}}
In this paper we propose a new task of multimedia event extraction and setup a new benchmark.
We also develop a novel multimedia structured common space construction method to take advantage of the existing image-caption pairs and single-modal annotated data for weakly supervised training. 
Experiments 
demonstrate its effectiveness 
as a new step towards semantic understanding of events in multimedia data. 
In the future, we aim to extend our framework to extract events from videos, and make it scalable to new event types. We plan to expand our annotations by including event types from other text event ontologies, as well as new event types not in existing text ontologies. We will also apply our extraction results to downstream applications including cross-media event inference, timeline generation, etc. 

\section*{Acknowledgement}
This research is based upon work supported in part by U.S. DARPA AIDA Program No. FA8750-18-2-0014 and U.S. DARPA KAIROS Program No. FA8750-19-2-1004. The views and conclusions contained herein are those of the authors and should not be interpreted as necessarily representing the official policies, either expressed or implied, of DARPA, or the U.S. Government. The U.S. Government is authorized to reproduce and distribute reprints for governmental purposes notwithstanding any copyright annotation therein. 

\bibliography{acl2019}
\bibliographystyle{acl_natbib}

\end{document}